\documentclass[a4paper,11pt]{article}
\usepackage{pos}

\title{IBPs and differential equations in parameter space}
\ShortTitle{Parameter-space Feynman integrals}

\author[a]{Daniele Artico}
\author*[b]{Lorenzo Magnea}

\affiliation[a]{Institut f\"ur Physik, Humboldt-Universit\"at zu Berlin, \\
Newtonstra\ss e 15, D-12489, Berlin, Germany}

\affiliation[b]{Dipartimento di Fisica, Universit\`a di Torino, 
and INFN, Sezione di Torino, \\ Via Pietro Giuria 1, I-10125 Torino, Italy}

\emailAdd{daniele.artico@physik.hu-berlin.de}
\emailAdd{lorenzo.magnea@unito.it}

\abstract{We present a projective framework for the construction of Integration
by Parts (IBP) identities and differential equations for Feynman integrals, working
in Feynman-parameter space. This framework originates with very early results 
which emerged long before modern techniques for loop calculations were 
developed~\cite{Regge:1968rhi,Ponzano:1969tk,Ponzano:1970ch,Regge:1972ns,
Barucchi:1973zm,Barucchi:1974bf}. Adapting and generalising these results
to the modern language, we use simple tools of projective geometry to 
generate sets of IBP identities and differential equations in parameter space, 
with a technique applicable to any loop order. We test the viability of the method 
on simple diagrams at one and two loops, providing a unified viewpoint 
on several existing results.}

\FullConference{16th International Symposium on Radiative Corrections: 
Applications of Quantum Field Theory to Phenomenology (RADCOR2023)\\
28th May - 2nd June, 2023\\
Crieff, Scotland, UK\\}

\newcommand{\be}{\begin{equation}}
\newcommand{\ee}{\end{equation}}
\newcommand{\beq}{\begin{eqnarray}}
\newcommand{\eeq}{\end{eqnarray}}
\newcommand{\bea}{\begin{eqnarray}}
\newcommand{\eea}{\end{eqnarray}}
\newtheorem{theorem}{Theorem}
\usepackage{tikz-feynman}
\newcommand{\secn}[1]{Section~\ref{#1}}
\def\eq#1{Eq.~(\ref{#1})}

\begin{document}
\maketitle

%%%%%%%%%%%%%%%%%%%%%%%%%%%%%%%%%%%%%%%%%%%%%

\section{Introduction}
\label{Intro}

The calculation of high-order Feynman integrals is crucial for the current and future 
precision physics programs at particle accelerators~\cite{Heinrich:2020ybq}: to this 
end, modern methods have evolved that go much beyond a direct evaluation. These 
developments began with the discovery of Integration-by-Parts (IBP) identities in 
dimensional regularization~\cite{Tkachov:1981wb,Chetyrkin:1981qh}, and the 
introduction of the method of differential equations~\cite{Kotikov:1990kg,Remiddi:1997ny,
Gehrmann:1999as}: calculations are now typically tackled with automatic algorithms 
combining these ideas~\cite{Laporta:2000dsw}. Further developments have involved an 
enhanced understanding of the role of generalised unitarity, and of the linear functional spaces 
where classes of Feynman integrals reside~\cite{Duhr:2011zq,Duhr:2019wtr,Abreu:2022mfk,
Britto:2023rig}, and an optimised use of dimensional regularization~\cite{Henn:2013pwa}. 
A vast amount of work along these lines has significantly expanded the range of processes 
for which high-order calculations are possible, and has much deepened our mathematical 
understanding of Feynman integrals~\cite{Heinrich:2020ybq,Weinzierl:2022eaz}.

Interestingly, the historical exploration of Feynman integrals via IBPs and differential 
equations predates these recent developments. Indeed, the projective nature of 
Feynman parameter integrands and the monodromy properties of Feynman integrals 
attracted early attention from mathematicians and physicists already in the late 
1960's~\cite{Eden:1966dnq,Pham,Lascoux,Regge:1968rhi}. Around that time, in particular,
Tullio Regge and his collaborators explored the monodromy properties of various 
classes of Feynman integrals~\cite{Regge:1968rhi,Ponzano:1969tk,Ponzano:1970ch,
Regge:1972ns}, offering several insights that align with (and predate) contemporary 
findings. For example Regge argued, already at the time~\cite{Regge:1968rhi}, that 
Feynman integrals belong to a class of generalised hypergeometric functions, and
proposed that these functions should satisfy differential equations of the Picard-Fuchs 
type.

Although actual computational algorithms didn't emerge from these studies, Barucchi 
and Ponzano~\cite{Barucchi:1973zm,Barucchi:1974bf} constructed an explicit implementation
of Regge's ideas, applicable to one-loop diagrams. The corresponding Feynman integrals, 
in parametric form, were organised into sets connected by difference equations, akin to 
currently used IBPs; linear systems of homogeneous differential equations in the 
Mandelstam invariants were then derived, mirroring the known one-loop monodromy 
structure.

This note, summarising the results presented in~\cite{Artico:2023jrc}, builds on the 
work of Regge and collaborators, to propose a projective framework for deriving IBP 
identities and systems of linear differential equations for Feynman integrals, directly 
in parameter space. The framework accommodates dimensional regularisation, 
extends to infrared-divergent integrals, and generalises naturally beyond on loop. 
Interestingly, our results also underscore the role of boundary terms in IBP identities 
within the projective framework: unlike the momentum-space approach in dimensional 
regularization, these terms do not generally vanish in parameter space, and indeed 
they play a critical role in connecting complex integrals to simpler ones.

We begin our note by setting up conventions for Feynman integrals in parameter form, 
in \secn{Notatio}. Next, in \secn{Proiectio}, we introduce projective forms, and we use 
their properties to show how one can construct systems of difference equations for 
generic projective integrals. In \secn{Integratio} we focus on Feynman integrals, and provide
a general procedure to construct IBPs in this context, developing the one-loop case 
in some detail as an example. Explicit one-loop examples are presented in \secn{Monulo}, 
and two-loop examples in \secn{Binulo}. Finally, \secn{Conclusio} briefly discusses 
perspectives for future work. 

%%%%%%%%%%%%%%%%%%%%%%%%%%%%%%%%%%%%%%%%%%%%%

\section{Notations}
\label{Notatio}

Scalar Feynman integrals arise in loop-level perturbative calculations in any quantum
filed theory. In a momentum space formulation they take the form
\beq
  I_G \left( \nu_i, d \right) \, = \, (\mu^2)^{\nu - l d/2} \int \prod_{r = 1}^{l} 
  \frac{d^d k_r}{{\rm i} \pi^{d/2}} \, \prod_{i = 1}^n 
  \frac{1}{\left( - q_i^2 + m_i^2 \right)^{\nu_i}} \, , \qquad q_i \, = \, 
  \sum_{r = 1}^l \alpha_{ir} k_r + \sum_{j = 1}^m \beta_{ij} p_j \, ,
\label{FeyInt}
\eeq
where $q_i$ are the momenta flowing in each propagator, $k_r$ are the independent
loop momenta, and $p_j$ are the external momenta, while $d$ is the space-time 
dimension, and the integer exponents $\nu_i$ satisfy $\sum_i \nu_i = \nu$. The 
integration over loop momenta in \eq{FeyInt} can be performed in full generality 
by means  of the Feynman parameter technique. Using the notations from 
Refs.~\cite{Bogner:2007mn,Bogner:2010kv,Weinzierl:2022eaz}, the integral 
becomes
\beq
\label{FP}
  I_G \left( \nu_i, d \right) \, = \, \frac{\Gamma(\nu - l d/2)}{\prod_{j = 1}^n 
  \Gamma(\nu_j)} \int_{z_j \geq 0} d^n z \, 
  \delta \left(1 - \sum_{j = 1}^n z_j \right) \,
  \left( \prod_{j = 1}^n z_j^{\nu_j -1}\right) \,
  \frac{\mathcal{U}^{\, \nu - (l+1) d/2}}{\mathcal{F}^{\, \nu - l d/2}} \, ,
\eeq
where the Symanzik polynomials $\cal{U}$ and ${\cal F}$,
\beq
\label{Sym pol}
  \mathcal{U} \, = \, \sum_{\mathcal{T}_G} \prod_{i \in \mathcal{T}_G} z_i \, , 
  \qquad \quad 
  \mathcal{F} \, = \, \sum_{\mathcal{C}_G} \frac{\hat{s} 
  \left( \mathcal{C}_G \right)}{\mu^2} \, 
  \prod_{i \in \mathcal{C}_G} z_i \, - \, \mathcal{U} \sum_{i \in \mathcal{I}_G} 
  \frac{m_i^2}{\mu^2} \, z_i \, ,
\eeq
can be defined purely from the graph properties. To this end, let us denote by 
$\mathcal{I}_G$ the set of the internal lines of $G$, each endowed with a 
Feynman parameter $z_i$. A \textit{co-tree} $\mathcal{T}_G \subset \mathcal{I}_G$ 
is a set of internal lines of $G$ such that that the lines in its complement 
$\overline{\mathcal{T}}_G \subset \mathcal{I}_G$ form a spanning tree. 
Similarly, consider subsets ${\cal C}_G \subset {\cal I}_G$ with the property 
that, upon omitting the lines of ${\cal C}_G$ from $G$, the graph becomes a
disjoint union of two connected subgraphs. Each subset ${\cal C}_G$ defines
a {\it cut} of graph $G$, and contains $l+1$ lines; an invariant mass $\hat{s} 
\left({\cal C}_G \right)$ can be associated with each cut, by squaring the sum 
of the momenta flowing in (or out) of one of the two subgraphs. The Symanzik 
polynomial $\mathcal{U}$ is homogeneous of degree $l$, while the Symanzik 
polynomial $\mathcal{F}$ is homogeneous of degree $l+1$, so that the 
integrand (measure included) is homogeneous of degree $0$.

%%%%%%%%%%%%%%%%%%%%%%%%%%%%%%%%%%%%%%%%%%%%%

\section{A projective framework}
\label{Proiectio}

A crucial mathematical property of Feynman integrals is that their integrands are 
projective forms in the space of Feynman parameters, which can be identified with
$\mathbb{PC}^{n-1}$. This property is crucial for the characterisation of the function
spaces to which the integrals belong, and to many techniques for their explicit 
evaluation. The relevance of projective invariance was understood since the 
earliest systematic studies of Feynman diagrams~\cite{Regge:1968rhi}. In this 
Section, we provide an extremely concise summary of the relevant ideas.

In order to introduce projective forms, begin by considering a generic subset $A$, 
$|A| = a$, of the set $D = \left\{ 1, \ldots, N \right\}$, and define the $a$-form
\beq
\label{wA}
  \omega_A \, = \, d z_{i_1} \wedge ... \wedge dz_{i_a} \, ,
\eeq
with $i_1 < \ldots < i_a$. One can show that $\omega_A$ integrates to the projective 
$(a-1)$-form
\beq
\label{eta}
  \eta_A \, = \, \sum_{i \in A} \, \epsilon_{i, A - i} \, z_i  \, \omega_{A - i} \, , \qquad
  d \eta_A \, = \, a \omega_A \, ,
\eeq
where we introduced the signature factor
\beq
\label{epsilon}
  \epsilon_{k,B} \, = \, (-1)^{|B_k|} \, ,  \qquad   B_k \, = \, \left\{ i \in B, i < k \right\} \, .
\eeq 
As an example, for $A = \{ 1,2,3 \}$ one finds
\beq
\label{eta example}
  \eta_{ \{1,2,3\} } \, = \, z_1 \, dz_2 \wedge dz_3 - z_2 \, dz_1 \wedge dz_3 + z_3 \, dz_1 
  \wedge dz_2 \, .
\eeq
Projective forms such as $\eta_A$ are homogeneous of degree 1 in each coordinate
$z_i$, and they can serve as measures of integration for projective integrals. Indeed,
parametric Feynman integrands can be represented in the general form
\beq
\label{ProjectiveForm}
  \alpha_{n-1} \, = \, \eta_{n-1} \,\, \frac{Q \big( \left\{ z_i \right\} 
  \big)}{D^P \big( \left\{ z_i \right\} \big)} \, , 
\eeq
where $D \big( \left\{ z_i \right\} \big)$ and $Q \big( \left\{ z_i \right\} \big)$ are polynomials
of degrees such that the form (measure included) is homogeneous of degree $0$. 
A well-known example is the integrand for the massless one-loop box integral, which 
reads
\beq
\label{BoxExample}
  \psi_3 \left( \lambda, r \right) \, = \,
  \frac{(z_1 + z_2 + z_3 + z_4)^{\lambda}}{\left(r \, z_1 z_3 + z_2 z_4 
  \right)^{2 + \lambda/2}} \, \,
  \eta_{ \left\{ 1, 2, 3, 4 \right\}} \, .
\eeq
For finite integrals both $P$ in \eq{ProjectiveForm} and $\nu$ in \eq{FP} are integers. 
In the presence of divergences, as is the case for the massless box, we can incorporate
dimensional regularisation by allowing for general values of $d$, and thus of $\lambda$
in \eq{BoxExample}.
 
Two theorems naturally emerge within this projective framework. First of all, an essential
property of projective forms is the following~\cite{Regge:1968rhi}.
\begin{theorem}
The boundary of a projective form is itself projective.
\end{theorem}
\noindent This theorem arises from the properties of the operator
\beq
\label{Defp}
  p \, : \,  \sum_{|A| = q} R_A (z_i) \, \omega_A \quad \rightarrow \quad
  \sum_{|A| = q} R_A (z_i) \, \eta_A \, ,
\eeq
mapping affine $q$-forms into projective $(q-1)$-forms. It can be shown to satisfy
\beq
p^2 = 0 \: , \qquad   d \circ p + p \circ d \, = \, 0 \, .
\eeq
Based on these properties, a proof of the theorem can be found in~\cite{Artico:2023jrc}.

Next, it is possible to show that $\alpha_{n-1}$ is a closed form, while $\eta_{n-1}$
 is null on any surface defined by $z_i = 0$. A second theorem then follows
\begin{theorem}
Given two integration domains, $O, O' \in \mathbb{C}^n$, if their image in 
$\mathbb{PC}^{n - 1}$ is the same simplex, then $\int_O \alpha_{n - 1} \, = \, 
\int_{O'} \alpha_{n-1}$.
\end{theorem}
\noindent This theorem, also known as Cheng-Wu theorem~\cite{Cheng:1987ga}, 
allows, in practice, to set to zero any subset of the $n$ parameters $z_i$ in the 
argument of the $\delta$ function in \eq{FP}, providing a useful tool for concrete 
calculations.

%%%%%%%%%%%%%%%%%%%%%%%%%%%%%%%%%%%%%%%%%%%%%

\section{Integration by parts in projective space}
\label{Integratio}

The correspondence between projective forms and parametric integrals is obtained 
from the usual choice of chart in projective space identifying a coordinate symplex 
in $\mathbb{R}^n$ with the choice $\sum_{i = 1}^n z_i = 1$. With this choice one
finds simply
\beq
\label{projective F}
  \int_{S_{n - 1}} \, \eta_{n - 1} \, \frac{Q(z)}{D^P(z)} \, = \, \int_{z_i \geq 0}
  d z_1 \ldots d z_n \, \, \delta \! \left(1 - \sum_{i=1}^n z_i \right) \, 
  \frac{Q(z)}{D^P(z)} \, .
\eeq
We now show how the projective structure just introduced allows to easily construct
sets of difference equations connecting families of Feynman integrals, which play 
the role of the conventional IBP identities usually derived in momentum space. 
To this end, consider the projective $(n-2)$-forms
\beq
\label{w}
  \omega_{n-2} \, \equiv \, \sum_{i  = 1}^{n} (-1)^i \, \eta_{\{z\} - z_i} \,
  \frac{H_i (z)}{(P - 1) \, \big( D(z) \big)^{P - 1}} \, ,
\eeq
where $\eta_{\{z\} - z_i}$ denotes the projective volume form in $\mathbb{PC}^{n - 2}$,
obtained by omitting the coordinate $z_i$, and $H_i(z)$ are polynomials with a degree
chosen (together with $P$) to ensure projectivity. Differentiating these forms generates
(at the integrand level) a set of identities among parametric integrals, which correspond
to those obtained via integration by parts. One finds 
\beq
\label{dw}
  d \omega_{n-2} \, = \, \frac{1}{(P-1) \, \big( D(z) \big)^{P - 1}} \, \, \eta_{\{z\}} \, 
  \sum_{i = 1}^n \frac{\partial H_i(z)}{\partial z_i} \, - \, \frac{\eta_{\{z\}}}{\big( D(z) 
  \big)^P} \, \sum_{i = 1}^n  H_i \, \frac{\partial D(z)}{\partial z_i} \, .
\eeq
\eq{dw} plays a central role in our method. By suitably choosing the polynomials
$H_i(z)$, it allows to close systems of linear differential equations for Feynman 
integrals that can be used to compute them, just as usually done in the momentum 
space approach. We emphasise that these identities apply for any number 
of loops or external legs. In the remainder of this section, we will discuss a 
concrete implementation at the one-loop level developing the ideas of 
Ref.~\cite{Barucchi:1973zm}.

At one loop, parametric integrals have the general form
\beq
\label{1LFP}
  I_G (\nu_i, d) \, = \, \frac{\Gamma(\nu - d/2)}{\prod_{j = 1}^n \Gamma(\nu_j)}
  \int_{z_j \geq 0} \! d^n z \, \delta \bigg(1 - z_{n+1} \bigg) \, 
  \frac{\prod_{j = 1}^{n+1} z_j^{\nu_j -1}}{\bigg[
  \sum_{i=1}^{n+1} \sum_{j=1}^{i-1} s_{ij} z_i z_j \bigg]^{\nu - d/2}} \, ,
\eeq
where we introduced the notations
\beq
  z_{n+1} \, \equiv \, \sum_{i =1}^n z_i \, , \qquad \quad \nu_{n+1} 
  \, \equiv \, \nu - d + 1 \, ,
\eeq
and for the Mandelstam invariants we use
\beq
\label{sij}
  s_{ij} \, = \, \frac{(q_j - q_i)^2}{\mu^2} \quad (i,j = 1, \ldots, n) \, , \qquad \quad
  s_{i, n+1} \, = \, s_{n+1, i} \, \equiv \, - \frac{m_i^2}{\mu^2} \, .
\eeq
We now make the simplest and natural choice in \eq{dw}, picking the polynomials
$H_i(z)$ to coincide with the numerator of the relevant integral, for each value of $i$.
Thus we pick
\beq
\label{chooseHi}
  H_i \, = \, \delta_{ih} \left( \prod_{j = 1}^n z_j^{\nu_j -1} \right) 
  \left(\sum_{k =1}^n z_k \right)^{\nu - d} 
  \, = \, \, \delta_{ih} \, \prod_{j = 1}^{n+1} z_j^{\nu_j -1} \, ,
 \eeq
for $h = 1, \ldots, n$. Applying this choice produces a one-loop `integration-by-parts' 
identity that can be written as follows~\cite{Barucchi:1973zm}:
\beq
\label{1LIBP2}
  d \omega_{n-2}  + \sum_{k = 1}^{n+1} (s_{k h} + s_{k, n+1}) \,
  f \big( \! \left\{ \mathcal{R} - k \right\}_0,\left\{ k \right\}_{1} \! \big) \! & = & \!
  \frac{\nu_h - 1}{\nu - (d+1)/2} \, f \big( \! \left\{ h \right\}_{-1}, 
  \left\{ \mathcal{R} - h \right\}_0 \! \big) \\
  \! & + & \! \frac{\nu - d}{\nu - (d+1)/2} \, f \big( \! \left\{ n+1 \right\}_{-1}, 
  \left\{ \mathcal{R} - \left\{ n+1 \right\} \right\}_0 \! \big) \, , \nonumber
\eeq
where, following Ref.~\cite{Barucchi:1973zm}, we introduced an index notation
such that, for the function
\beq
\label{notforf1}
  f \big( \! \left\{ \nu_1, \ldots, \nu_{n+1} \right\} \! \big) \, \equiv \, 
  f \big( \! \left\{ {\cal R} \right\} \! \big) \, = \, \eta_{\{z\}} \, \frac{
  \prod_{j = 1}^{n+1} z_j^{\nu_j -1}}{\left( 
  \sum_{i = 1}^{n+1} \sum_{j=1}^{i-1} s_{ij} z_i z_j \right)^{\nu - d/2 }} \, ,
\eeq
we can raise or lower the exponents $\nu_i$ by adding $\{-1, 0,1\}$ in the subsets 
$\cal I$, $\cal J$ and $\cal K$ of set ${\cal R}$ respectively, and we denote the 
resulting function by
\beq
\label{notforf2}
  f \big( \! \left\{ {\cal I} \right\}_{-1}, \left\{ {\cal J} \right\}_{0}, 
  \left\{ {\cal K} \right\}_{1} \! \big) \, .
\eeq
Note that the exponent of the denominator in \eq{notforf1} is adjusted accordingly, 
to maintain projective properties. Note also that the action of raising and lowering
exponents according to the convention in \eq{notforf2} is subject to a constraint,
arising from the definition of $\mathcal{U}$ in \eq{Sym pol}. Specifically, the 
following sum rule holds
\beq
\label{sum of f}
  \sum_{i = 1}^n f \big( \! \left\{ \mathcal{R} - i \right\}_0, \left\{ i \right\}_{1} \! \big) 
  \, = \,  
  f \big( \! \left\{ \mathcal{R} - \left\{ n+1 \right\} \right\}_0, \left\{ n+1 \right\}_{1} 
  \! \big) \, .
\eeq
As we will see in the next section, \eq{1LIBP2} and \eq{sum of f} can be used to 
close systems of differential equations, leading to the determination of the one-loop 
Feynman integrals under study. Two-loop examples will be discussed in \secn{Binulo}.

%%%%%%%%%%%%%%%%%%%%%%%%%%%%%%%%%%%%%%%%%%%%%

\section{One-loop examples}
\label{Monulo}

In this section, we present two explicit examples of the use of \eq{1LIBP2}. Consider 
first the massless one-loop box integral, setting $t/s \equiv r$ and with all momenta 
incoming. Using dimensional regularisation, we define
\beq
\label{massless box}
  I_{\textrm{box}} \, \equiv \, \Gamma (2 + \epsilon) \, \int_{S_{n-1}} \eta_{\{ z \}} \,
  \frac{ (z_1 + z_2 + z_3 + z_4)^{2 \epsilon}}{\left( r  z_1 z_3 + z_2 z_4 
  \right)^{2 + \epsilon}} \, \equiv \, \Gamma( 2 + \epsilon) \, I(1,1,1,1; 2 \epsilon) \, ,
\eeq
where for the box family of integrals we use the notation $I(\nu_1, \nu_2, \nu_3 , 
\nu_4; \nu_5)$. Differentiating with respect to $r$ raises two indices by one unit, 
as in	
\bea
\label{dzI1}
  &\partial_r I (1, 1, 1, 1; 2 \epsilon) \, = \, - (2 + \epsilon) \, I(2, 1, 2, 1; 2 \epsilon) \, \\
    &\partial_r  I (2, 1, 2, 1; 2 \epsilon) \, = \, - ( 3 + \epsilon) \, I (3, 1, 3, 1; 2 \epsilon) \, .
\eea
According to a theorem by Barucchi and Ponzano~\cite{Barucchi:1973zm}, for
any one-loop diagram a system of differential equation can be set up, involving 
the desired integral, plus the ones obtained by lifting an even number of propagators 
by 1. For the massless box, following this construction we find that a closed system 
of differential equations can be obtained for the integrals\footnote{It is well-known
that a basis of master integrals for the massless box requires only three integrals. 
Here we are simply illustrating the Barucchi-Ponzano construction, which in this 
case yields an over-complete basis, and we have not attempted optimisations. 
On the other hand, the method correctly predicts the size of the basis for the 
most general one-loop diagram, as recently confirmed by Refs.~\cite{Bitoun:2017nre,
Bitoun:2018afx,Mizera:2021icv}.}
\beq
\label{basisbox}
  \Big\{ I(1, 1, 1, 1; 2 \epsilon), I(2, 1, 2, 1; 2 \epsilon), I(1, 2, 1, 2; 2 \epsilon),
  I(2, 2, 2, 2; 2 \epsilon) \Big\} \,.
\eeq
The system is obtained by using identities generated by \eq{1LIBP2}, such as, 
for example,
\beq
\label{2010}
  r I(3, 1, 3, 1; 2 \epsilon) + \int d \omega_{n-2} \, = \, 
  \frac{2}{3 + \epsilon} I(2, 1, 2, 1; 2 \epsilon) + \frac{2 \epsilon}{3 + \epsilon} 
  I(3, 1, 3, 1; - 1 + 2 \epsilon) \, .                                                                                                                                                                                                                                                                                                                                                                                
\eeq
where the integral of $d \omega_{n-2}$ gives a vanishing boundary term, since
\beq
\label{vanbou}
  \frac{z_1^2 z_3 
  \left(z_1 + z_2 + z_3 + z_4 \right)^{2 \epsilon}}{(3 + \epsilon)
  (r z_1 z_3 + z_2 z_4)^{3 + \epsilon}}
  \big( z_2 d z_3 \wedge dz_4 - z_3 d z_2 \wedge d z_4 + z_4 
  d z_2 \wedge d z_3 \big) \Bigg|_{\partial S_{n-1}}
  \, = \, 0 \, .
\eeq
The system of differential equations obtained in this way can be written as
\beq
\label{system}
  \partial_r \textbf{b} \, \equiv \, \partial_r \left(
  \begin{array}{c}
  I(1, 1, 1, 1; 2 \epsilon) \\
  I(2, 1, 2, 1; 2 \epsilon) \\
  I(1, 2, 1, 2; 2 \epsilon) \\ 
  I(2, 2, 2, 2; 2 \epsilon)
  \end{array}
  \right) =  \left(
  \begin{array}{c c c c}
  0 & - (2 + \epsilon) & 0 & 0 \\
  0 & - \frac{3 + \epsilon}{r} & 0 & - \frac{3  + \epsilon}{r} \\
  0 & 0 & 0 & - (3 + \epsilon) \\
  0 & - \frac{1}{(3 + \epsilon) r ( 1 + r)} & \frac{1}{(3 + \epsilon) r (1 + r)} & - 
  \frac{1 + \epsilon + 3 r}{(3 + \epsilon) r (1 + r)} 
  \end{array}
  \right)
  \textbf{b} \, .
\eeq
This system can be brought to canonical form by using (for example) the technique
of Magnus exponentiation \cite{Magnus:1954zz}. It can then be solved by iteration, 
and the solution, consistently with the literature~\cite{Henn:2014qga}, is given by
\beq
  I_{\rm box} & = & \frac{k(\epsilon)}{r} \bigg[ \frac{1}{\epsilon^2} - \frac{\log r}{2 \epsilon} -
  \frac{\pi ^2}{4} + \epsilon \, \bigg( \frac{1}{2} \, {\rm Li}_3 (- r) - \frac{1}{2} \, {\rm Li}_2 (- r) 
  \log r + \frac{1}{12} \log ^3 r \nonumber \\
  && \hspace{2cm} - \, \frac{1}{4} \log(1 + r) \left(\log ^2 r + \pi ^2 \right) 
  + \frac{1}{4} \, \pi ^2 \log r + \frac{1}{2} \zeta(3) \bigg) \, + \, {\cal O} (\epsilon^2)
  \bigg] \, ,
\eeq
with $k(\epsilon) \, = \, 4-\frac{\pi^2}{3}\epsilon^2 -\frac{40 \zeta(3)}{3}\epsilon^3$.

The difference equations generated in parameter space by \eq{1LIBP2} effectively
include also dimensional-shift identities, and they connect the desired integrals to
lower-point integrals through non-vanishing boundary terms. As an example, 
consider the following identity for five-point integrals:
\beq
\label{I00000}
  \int_{S_{\left\lbrace 1,2,3,4,5 \right\rbrace}} \! d \omega_3 + s_{13} \, 
  I(1, 1, 2, 1, 1; 2 \epsilon) + s_{14} \, I(1, 1, 1, 2, 1; 2 \epsilon) 
  \, = \, \frac{2 \epsilon}{2 + \epsilon} I(1, 1, 1, 1, 1; - 1 + 2 \epsilon) \, , \qquad
\eeq
with 
\beq
  d \omega_3 \, = \, d \left[ - \, \eta_{\{2,3,4,5\}} \, 
  \frac{(z_1 + z_2 + z_3 + z_4 + z_5)^{2 \epsilon}}{(2 + \epsilon)
  \left( s_{13} z_1 z_3 + s_{14} z_1 z_4 + s_{24} z_2 z_4 + s_{25} z_2 z_5 + 
  s_{35} z_3 z_5 \right)^{2 + \epsilon}} \right] \, .
\eeq
The integration of this form using Stokes theorem produces a non-vanishing 
boundary term, corresponding to the one-loop box integral with one external 
leg off-shell. Specifically, one finds
\beq
  \int_{S_{\{2,3,4,5\}}} \! \eta_{\{2,3,4,5\}} \, 
  \frac{(z_2 + z_3 + z_4 + z_5)^{2 \epsilon}}{\left(s_{24} z_2 z_4 + s_{25} z_2 z_5 + 
  s_{35} z_3 z_5 \right)^{2 + \epsilon}} \, = \,  
  I_{\rm box}^{(1)} (s_{25}) \, ,
\eeq
where in this case $s_{25}$ is the mass of the off-shell leg. Using similar 
identities, dimensional-shift relations for the one-loop pentagon~\cite{Bern:1993kr} 
can easily be reproduced. One finds
\beq
\label{pentagon}
  2 (2 + \epsilon) \,
  I(1, 1, 1, 1, 1;1 + 2 \epsilon) & = & 
  \Bigg\{\frac{s_{13} s_{24}-s_{13} s_{25}-s_{14} s_{25}+s_{14} s_{35}-
  s_{24} s_{35}}{s_{13} s_{14} s_{25}} \, I_{\rm box}^{(1)} (s_{25}) \nonumber \\ 
  && - \, \frac{s_{13} s_{24}+s_{13} s_{25}-s_{14} s_{25}+s_{14} s_{35}-
  s_{24} s_{35}}{s_{13} s_{24} s_{25}} \, I_{\rm box}^{(2)} (s_{13}) \nonumber \\ 
  && - \, \frac{s_{13} s_{24}-s_{13} s_{25}+s_{14}
   s_{25}-s_{14} s_{35}+s_{24} s_{35}}{s_{13} s_{24} s_{35}} \, I_{\rm box}^{(3)} (s_{24}) \nonumber \\
  && + \, \frac{s_{13} s_{24}-s_{13} s_{25}+s_{14} s_{25}-s_{14} s_{35}-s_{24}
   s_{35}}{s_{14} s_{24} s_{35}} \, I_{\rm box}^{(4)}(s_{35}) \nonumber \\ 
  && - \, \frac{s_{13} s_{24}-s_{13} s_{25}+s_{14} s_{25}+s_{14} s_{35}-
  s_{24} s_{35})}{s_{14} s_{25} s_{35}} \, I_{\rm box}^{(5)} (s_{14}) \Bigg\} \nonumber \\ 
  && + \, 2 \epsilon \, I(1, 1, 1, 1, 1; - 1 + 2 \epsilon) \, .
\eeq
Since the integral in the last line is finite in $d=4$, this gives the well-known result 
stating that the massless pentagon integral is given by a liner combination of box
integrals, up to corrections vanishing in four dimensions.

%%%%%%%%%%%%%%%%%%%%%%%%%%%%%%%%%%%%%%%%%%%%%

\section{Two-loop examples}
\label{Binulo}

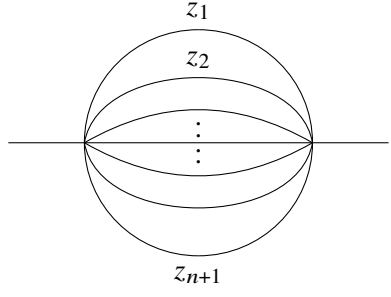
\begin{figure}[!h]
\centering
\begin{tikzpicture}
            \begin{feynman}
            \vertex (a1);
            \vertex[right=1cm of a1] (a2);
            \vertex[above right= 2.12 cm of a2](a3);
            \vertex[below right= 2.12 cm of a2](a4);
            \vertex[below = 1 cm of a3] (a31);
            \vertex[above = 1 cm of a4] (a32);
            \vertex[right=3cm of a2] (a5); 
            \vertex[right=1cm of a5] (a6); 
            \diagram* {
            (a1) -- (a2)
                -- [quarter left] (a3)
                -- [quarter left] (a5)
                -- [quarter left] (a4)
                -- [quarter left] (a2),
                (a5) -- (a6),
                (a2) -- (a5)
            };
            \end{feynman}
            \draw (a2) to[out=30,in=150] (a5);
            \draw (a2) to[out=-30,in=-150] (a5);
            \draw (a2) to[out=80,in=100] (a5);
            \draw (a2) to[out=-80,in=-100] (a5);
            \path
            (2.5,1.75) node {$z_1$}
            (2.5,1.1) node {$z_2$}
            (2.5,0.25) node {.}
            (2.5,0.10) node {.}
            (2.5,-0.25) node {.}
            (2.5,-0.10) node {.}
            (2.5,-1.75) node {$z_{n+1}$};
            \end{tikzpicture}
\caption{\label{banana} Sunrise diagram at $l$ loops.}
\end{figure}

We now very briefly discuss the application of the method beyond one loop. A first interesting
case is given by the family of $l$-loop sunrise diagrams, {\it i.e.} diagrams  contributing to 
a two-point function, with two vertices connected by $l+1$ propagators, illustrated in 
Fig.~\ref{banana}. These integrals have been extensively studied in recent years, since 
they provide a natural laboratory for multi-loop calculation, and in particular, with massive 
legs, provide the simplest example of integrals involving elliptic curves, and thus yielding 
functions beyond polylogarithms (see, for example,~\cite{Broadhurst:1993mw,Muller-Stach:2011qkg,
Adams:2013nia,Bloch:2013tra,Adams:2015ydq,Ablinger:2017bjx,Bogner:2019lfa,Bourjaily:2018yfy,
Broedel:2021zij,Bonisch:2021yfw,Bourjaily:2022bwx} and references therein).

The first Symanzik polynomial for $l$-loop sunrise integrals is given by
\beq
  \mathcal{U}_l \, = \, \sum_{i = 1}^{l+1} z_1 \ldots \hat{z_i} \ldots z_{l+1} \, ,
\label{banfirstsym}
\eeq
where $\hat{z_i}$ denotes the omission of $z_i$. \eq{banfirstsym} displays the high degree 
of symmetry of the graph, while the second Symanzik polynomial ${\cal F}$ depends
on the configuration of masses on the internal legs.
In the specific case of $l = 2$ and equal internal masses, the Feynman parametric integral is
\beq
\label{sunrise}
  I \big( \nu_1, \nu_2 , \nu_3 ; \lambda_4 \big) \, = \! \int_{S_{\{1,2,3\}}}
  \frac{\eta_3 \, z_1^{\nu_1 - 1} z_2^{\nu_2 - 1} z_3^{\nu_3 - 1} \, 
  \big( z_1 z_2 +  z_2 z_3 + z_3 z_1)^{\lambda_4}}{\Big[ r\,  z_1 z_2 z_3 - 
  (z_1 + z_2 + z_3) \big( z_1 z_2 + z_2 z_3 + z_3 z_1 \big) 
  \Big]^{\frac{2 \lambda_4 + \nu}{3}}} \quad .
\eeq
By choosing a suitable numerator in our master identity, \eq{dw},
\beq
  H (z) \, = \, z_1^{\nu_1 - 1} z_2^{\nu_2 - 1} z_3^{\nu_3 - 1} 
  \big(z_1 z_2 + z_2 z_3 + z_3 z_1 \big)^{\lambda_4} \, ,
\label{Hsunrise}\
\eeq 
one easily derives integration by parts identities, and one can build a linear system
of differential equations that closes (as expected) on the three master integrals,
$I(1,1,1; 3 \epsilon)$, $I(2,1,1; 1 + 3 \epsilon)$, and the one-loop tadpole integral 
$I(2,2; 1 + 3 \epsilon)$. Interestingly, also in this case the non-vanishing boundary 
term provides an inhomogeneous contribution to the system. It arises form the 
basis integral
\beq
  \int d \omega_1 \, = \, \frac{1}{2 (1 + \epsilon)} \int_{S_{\{1,2\}}} \! \eta_{\{1,2\}} \,
  \frac{(z_1 z_2)^{\epsilon}}{\big[ - (z_1 + z_2) \big]^{2 + 2\epsilon}} \, = \, 
  \frac{(-1)^{2 \epsilon}}{2 + 2 \epsilon} \, 
  \frac{\Gamma^2(1 + \epsilon)}{\Gamma(2 + 2 \epsilon)} \, ,
\label{bouter}
\eeq 
corresponding to the massive one-loop tadpole. In two space-time dimensions, the 
sunrise integral is finite and the linear system can be analysed for $\epsilon = 0$. More 
precisely, as discussed in more detail in~\cite{Artico:2023jrc}, the first-order differential 
equations can be combined into a single second-order equation for the equal-mass 
sunrise, which has long been known to be of elliptic type~\cite{Broadhurst:1993mw,
Laporta:2004rb,Lairez:2022zkj}. We find
\beq
\label{elliptic}
  && \frac{r}{3} \, \frac{d^2}{d r^2} I(1, 1, 1; 0) \, + \, 
  \left( \frac{1}{3} + \frac{3}{r - 9} + \frac{1}{3 (r - 1)} \right) \frac{d}{d r} I(1, 1, 1; 0) 
  \nonumber \\
  && \hspace{2.6cm} \, - \, \left( \frac{1}{4 (r - 9)} + \frac{1}{12 (r - 1)} \right) I(1, 1, 1; 0) \, = \, 
  \frac{2}{(r - 1)(r - 9)} \, .
\eeq
It is important to note that the procedure we followed is not expected to generalise 
smoothly to the two-loop sunrise diagram with unequal masses, since a straightforward 
application of Stokes' theorem in that case must take into account the presence of 
singularities at the simplex boundaries: the difference between the two cases is 
discussed in detail in Ref.~\cite{Weinzierl:2020xyy}. We leave the analysis of the 
general case to future work. On the other hand, we note that our method readily
reproduces the classic results of Ref.~\cite{Chetyrkin:1981qh} for two-point, 
five-propagator integrals, which can be systematically reduced to four-propagator 
integrals yielding simple combinations of $\Gamma$ functions. Once again, boundary
terms play a distinctive role in parameter space, as discussed in detail in~\cite{Artico:2023jrc}.

%%%%%%%%%%%%%%%%%%%%%%%%%%%%%%%%%%%%%%%%%%%%%

\section{Perspectives}
\label{Conclusio}

In this note, we have summarised the results of Ref.~\cite{Artico:2023jrc}, where we 
introduced a projective framework for deriving IBP identities and differential equations 
for Feynman integrals directly in parameter space, building upon very early work by 
Tullio Regge and collaborators~\cite{Regge:1968rhi,Ponzano:1969tk,Ponzano:1970ch,
Regge:1972ns,Barucchi:1973zm,Barucchi:1974bf}. Specifically, we showed how these
early techniques can be adapted to include dimensional regularization, and how they 
can be generalised beyond one loop.

Comparing the parameter-space method to momentum-space approaches, it's clear 
that the organisation of calculations differs significantly. The integral bases and the 
resulting differential equations generated by the projective framework are in general 
distinct from the conventional ones. One notable aspect of this framework is the role 
played by boundary terms, which vanish in the momentum-space approach. In this case,
instead, they play a crucial role, linking complex integrals to simpler ones. We note also 
that parameter-space integrands closely mirror the graph symmetries, and circumvent issues 
related to loop-momentum routing and irreducible numerators, which can complicate 
momentum-space algorithms. Importantly, the projective framework aligns closely 
with the algebraic structures underpinning Feynman integrals, which may provide
direction for future progress.

The present work is largely a feasibility study: for the future, the goal is clearly to 
extend these techniques to more complex integrals, including higher-loop and multi-scale 
examples, possibly developing automated tools. Besides the obvious interest in direct 
physics applications, this will allow for a necessary detailed comparison of parameter-space 
and momentum-space approaches, including computational aspects.

%%%%%%%%%%%%%%%%%%%%%%%%%%%%%%%%%%%%%%%%%%%%%

\section*{Acknowledgments}

\noindent Research supported in part by the Italian Ministry of University and Research (MIUR), under 
grant PRIN 20172LNEEZ. DA is funded by the Deutsche Forschungsgemeinschaft (DFG, 
German Research Foundation) - Projektnummer 417533893/GRK2575 "Rethinking Quantum 
Field Theory". We thank {\tt ChatGPT} for efficiently summarising some of our lengthy 
arguments, leading to a more compact note.

%%%%%%%%%%%%%%%%%%%%%%%%%%%%%%%%%%%%%%%%%%%%%

%%%%%%%%%%%%%%%%%%%%%%%%%%%%%%%%%%%%%%%%%%%%%

%%%%%%%%%%%%%%%%%%%%%%%%%%%%%%%%%%

\end{document}